\documentclass[12pt]{article}
\usepackage{latexsym}
\usepackage{latexsym}
\usepackage{epsfig}
\usepackage{times}
\usepackage{amsfonts} % Use $\mathbb{R}$ to indicate the reals
\usepackage{float}
\def\mymedskip{\vskip\medskipamount}
\def\mymedbreak{\par \ifdim\lastskip<\medskipamount
  \removelastskip \penalty-100 \mymedskip \fi}
\def\myaftermedspace{\par \ifdim\lastskip<\medskipamount
  \removelastskip \penalty55\mymedskip\fi}
\newcommand{\eop}{{\unskip\nobreak\hfil\penalty50
          \hskip2em\hbox{}\nobreak\hfil$\Box$
          \parfillskip=0pt \finalhyphendemerits=0 \par}}
\newenvironment{proof}%
{\mymedbreak{\noindent\bf Proof:\enspace}}{\eop\myaftermedspace}
{\mymedbreak{\noindent\bf Proof of Theorem #1:\enspace}}{\eop\myaftermedspace}
{\mymedbreak\noindent{\bf Remark:}%
\enspace\rm}%
{\myaftermedspace}
\newtheorem{teor}{Theorem}[section]
\newtheorem{examp}[teor]{Example}
\newtheorem{con}[teor]{Conjecture}
\newtheorem{prop}[teor]{Proposition}
\newtheorem{theo}[teor]{Theorem}
\newtheorem{cor}[teor]{Corollary}
\newtheorem{defi}[teor]{Definition}
\newtheorem{lem}[teor]{Lemma}
\newcommand{\beq}{\begin{equation}}
\newcommand{\eeq}{\end{equation}}
\newcommand{\beql}[1]{\begin{equation} \label{#1}}
\newcommand{\eeql}{\end{equation}}
\newcommand{\beqa}{\begin{eqnarray*}}
\newcommand{\eeqa}{\end{eqnarray*}}
\newcommand{\beqal}[1]{\begin{eqnarray} \label{#1}}
\newcommand{\eeqal}{\end{eqnarray}}
\newcommand{\beqan}{\begin{eqnarray}}
\newcommand{\eeqan}{\end{eqnarray}}
\newcommand{\bpf}{\begin{proof}}
\newcommand{\epf}{\end{proof}}
\newcommand{\bF}{{\bf F}}
\newcommand{\cA}{{\cal A}}
\newcommand{\cB}{{\cal B}}
\newcommand{\cH}{{\cal H}}
\newcommand{\supp}{{\rm supp}}
\newcommand{\bfh}{{\bf h}}
\newcommand{\bfx}{{\bf x}}

\newcommand{\wt}{{\rm wt}}

\begin{document}

\begin{titlepage}
\title{On parity check collections for iterative erasure decoding
that correct all correctable erasure patterns of a given size}
\author{Henk D.L.\ Hollmann and Ludo M.G.M.\ Tolhuizen
\thanks{The authors are with Philips Research Laboratories, Prof.\ Holstlaan 4,
5656 AA Eindhoven, The Netherlands;
e-mail:\{henk.d.l.hollmann,ludo.tolhuizen\}@philips.com}}
\maketitle
\begin{abstract}
\noindent
Recently there has been interest in the construction of small parity check sets
%added
for iterative decoding of
%end added
 the Hamming code with the property
that each uncorrectable (or stopping) set of size three is the
support of a codeword and hence uncorrectable anyway. Here we
reformulate and generalise the problem
%to erasure sets of size $m$ and improve on the known construction for $m=3$.
and improve on this construction.

%HIER ZOU IK EIGENLIJK WILLEN STOPPEN MAAR ANDERS KAN DIT ER NOG ACHTER:

First we show that a parity check collection
%LTv5
that corrects all correctable erasure patterns of size $m$ for the
$r$-th order Hamming code (i.e, the Hamming code with codimension
$r$)
%LTv5
%HH: Ik ben akkoord. Bij het nauwkeurig lezen viel me trouwens wel
%op dat dit niet is wat we doen, in feite!
%Namelijk: ...corrects one error from ... is ons ding...
 provides for {\em all\/} codes of codimension~$r$ a
corresponding ``generic'' parity check collection
%{\em all\/} codes of codimension $r$ which has
with this property. This leads naturally to a necessary and
sufficient condition on such generic parity check collections. We
use this condition to construct a generic parity check collection
for codes of codimension $r$ correcting all correctable erasure
patterns of size at most $m$, for all $r$ and $m\leq r$, thus
generalising the known construction for $m=3$. Then we discuss
optimality of our construction and show that it can be improved
for $m\geq 3$ and $r$ large enough. Finally we discuss some
directions for further research.
\end{abstract}
\thispagestyle{empty}
\end{titlepage}
\section{Introduction}
%
%HH: Alle \subset vervangen door \subseteq
%
This note addresses iterative decoding of erasures for a binary
linear code using a {\em given, fixed\/} collection of parity
check equations. The idea is to correct erasures in a codeword
one-by-one, where in each step a parity check equation is used
that involves {\em precisely one\/} of the remaining erasure
positions, thus allowing this erasure to be corrected. The
correction procedure stops if no such parity check can be found
for the set of current erasures; in that case the set of the
positions of these erasures is called a {\em stopping set\/} for
the given collection of parity checks
%LT
\cite{Di02}.
 As shown in \cite[Lemma 1.1]{Di02}, the correction procedure
stops with erasures in the positions of the largest stopping set
contained in the set of erased positions that we started with.

Each
%(full-rank)
subset of the dual code %(without the all-zero word)
can be used as collection of parity checks for this method.
(Mostly we will consider only full-rank subsets that do not
contain the all-zero word.)
%We remark that
Different subsets in general lead to different stopping sets. Note
however that the support of each nonzero codeword is always a
stopping set: indeed, by definition each parity check involves an
even number of positions from such a set.

%The erasure sets for which this iterative decoding procedure fails
%(i.e.\, eventually gets stuck)
%somewhere)
%are called {\em dead-end sets\/} \cite{WeAb05b}. It is not
%difficult to see that a set is dead-end precisely when it contains a stopping set:
%indeed, the erasures contained in a stopping set that is included in the erasure set
%are obviously non-removable.
A received word containing only correct symbols and erasures can
be decoded unambiguously precisely when exactly one codeword
agrees with this word in the non-erased positions; if the code is
linear this is the case precisely when no support of a nonzero
codeword is contained in the set of erasures. For this reason we
will refer to a set of erasure as {\em uncorrectable\/} if it
contains the support of a nonzero codeword, and as {\em
correctable\/} otherwise. It can be seen \cite[Thm. 8]{WeAb05b}
that the iterative algorithm decodes each correctable set of
erasures if the collection of parity checks consists of the entire
dual code.\footnote{In fact, it is shown in \cite[Thm.8]{WeAb05b}
that for a code of codimension $r$, this property holds if we take
the entire dual code without the all-zero word and $r-1$ arbitrary
other codewords.
%stopping set is
%in fact the disjoint union of supports of non-zero codewords.
% LT Naar footnote verhuisd en korter gemaakt
In Section~4 we will prove that this property even holds if we
take as parity check collection the complement of an
$(r-1)$-dimensional subspace of the dual code, and that this is
best possible for the Hamming code.}

Motivated by these observations, we refer
%to sets containing the
%support of a nonzero codeword (that is, uncorrectable erasure
%sets) as {\em unavoidable\/} dead-end sets and to other dead-end
%sets as {\em avoidable\/}. Note that a set of size one or two is a
%stopping set precisely when each parity check involves zero or two
%positions from the set; so if the parity checks span the dual
%code, then all dead-end sets of size one or two are uncorrectable,
%%or contain the support of a nonzero codeword,
%and hence are unavoidable.
to a parity check collection as {\em $m$-erasure reducing\/} if
each stopping set of size $m$ for this parity check collection is
uncorrectable. In other words, a parity check collection is
$m$-erasure reducing precisely if for any correctable pattern of
$m$ erasures, a parity check equation from our collection can be
used to remove a single erasure. We call a parity check collection
 {\em $m$-erasure correcting\/} if iterative decoding allows to decode {\em
 all} correctable patterns of $m$ erasures.

The design of a parity check collection to be used for such an
iterative decoding procedure involves a trade-off between the {\em
complexity\/} of the resulting decoding method, which is
determined amongst others by the {\em size\/} of the collection,
and the {\em effectiveness\/} of the method, which in the case of
a small erasure probability is mainly determined by the minimum
size of a stopping set and the number of stopping sets of this
size. From the above discussion we see that this minimum size can
be as large as the minimum distance~$d$ of the code, with all
stopping sets of size $d$ being supports of codewords, and
therefore it is interesting to investigate
%LT Andere formulering
$d$-erasure correcting
parity check collections. In \cite{WeAb05a}, which in fact
inspired the present work, this problem was investigated for the
$r$-th order $[n=2^r-1,k=2^r-r-1,d=3]$ Hamming codes. In that
paper, Weber and Abdel-Ghaffar constructed a $3$-erasure
correcting parity check collection of size
%$C(r-1,0)+C(r-1,1)+C(r-1,2)=
$1+r(r-1)/2$ for the $r$-th order Hamming code.
%LT Nee, dat vroegen ze niet. and asked whether that is optimal.

Our aim here is both to generalise and to improve this result. So
after Section~\ref{Sec:formal}, which contains some notation and
definitions, we start in Section~\ref{Sec:prop} with a
reformulation of the problem, in the following way. A collection
of parity checks $\cH$ for an $r$-th order Hamming code is fully
described by a fixed parity check matrix $H$ for this code
together with a specification of which linear combinations of the
$r$ rows of $H$ are contained in our collection $\cH$. Such a
specification essentially consists of a subset $\cA$ of $\bF_2^r$
describing these linear combinations, so that $\cH$ consists of
all parity checks $aH$ with $a$ in $\cA$. A not too difficult but
important insight is that if a certain collection $\cA\subseteq
\bF_2^r$ describes the linear combinations of an $m$-erasure
reducing (or correcting) parity check collection for the $r$-th
order Hamming code, then these linear combinations generate
$m$-erasure reducing (or correcting) parity check collections for
{\em all\/} codes with codimension $r$. We will refer to such
collections $\cA$ as {\em generic $(r,m)$-erasure reducing (or
correcting)\/} sets. Note that this result can be interpreted as
saying that
%LTv5 :r-th order weggelaten
Hamming codes are in a sense {\em the most
difficult\/} codes to design an $m$-erasure reducing or correcting
parity check collection for.

The above insight also leads in a natural way to a useful necessary and sufficient
condition for collections $\cA\subseteq \bF_2^r$ to
%LTv5
%generate $m$-erasure reducing
%parity check collections for codes with codimension $r$.
be generic $(r,m)$-erasure reducing. We will use this condition to
show that the distinction between ``reducing'' and ``correcting''
need not to be made. Indeed, we will show that if $\cA$ is generic
$(r,m)$-erasure reducing, then it is also generic $(r,m')$-erasure
reducing for all $m'\leq m$, and hence generic $(r,m)$-erasure
correcting. We will also give an example showing that a similar
property need not hold for
%{\em every\/}
an $m$-erasure reducing parity check collection for a {\em
specific\/} code.

In Section~\ref{Sec:constr}
we use the condition referred to above to obtain generic $(r,m)$-erasure
correcting sets of size
% LT Ik vind die C toch lelijk.
%$\sum_{i=0}^{m-1}C(r-1,i)$
%for all $m$, where $C(n,k)$ denotes the $k$-th $n$-th order binomial coefficient.
$\sum_{i=0}^{m-1}{r-1\choose i}$. For $m=3$ this construction
produces the $3$-erasure correcting parity check collection from
\cite{WeAb05a}.

Various optimality results are obtained in
Section~\ref{Sec:optimal}. We show that the construction from
Section~\ref{Sec:constr} is optimal for $m=r$ (and we conjecture
that it is also optimal for $m=r-1$) but not optimal for $m\geq3$
and large enough $r$. In particular, we show that the
construction for $m=3$ from \cite{WeAb05a} can be improved for
$r\geq5$.

Finally, in Section~\ref{Sec:claims}
we discuss our results, we indicate further directions of
research, and announce some further work on this and related problems.
%It turns out that for $m=3$ it is possible to
%
%
%LTv5 non-binary hier toegevoegd

%HH
We remark that most of the results of this paper can readily be
generalised to the non-binary case.
\section{Notations and definitions}\label{Sec:formal}
In this section, we introduce some notations and definitions.
Throughout this paper, we use boldface letters to denote row
vectors. All vectors and matrices are binary. If there is no
confusion about the length of vectors, we denote with ${\bf 0}$ and ${\bf 1}$
the vectors consisting of only zeroes or only ones, and with ${\bf e}_i$ the
$i$-th unit vector, the vector that has a one in position $i$ and zeroes
elsewhere.

The size of a set $A$ is  denoted by $|A|$.  If $H$ is a $r\times
n$ matrix and $E\subseteq\{1,2,\ldots ,n\}$, then the {\em restriction\/ $H(E)$
of $H$ to $E$\/} denotes
the $r\times |E|$ matrix consisting of those columns of $H$ indexed
by $E$. Similarly, if ${\bf x}\in\mathbb{F}_2^n$ and
$E\subseteq\{1,2,\ldots ,n\}$, then the restriction ${\bf x}(E)$ of ${\bf x}$ to $E$
is the vector of length $|E|$ consisting of the entries indexed by $E$.

%For a vector ${\bf x}\in \mathbb{F}_2^n$,
The {\em support\/} supp({\bf x}) of a vector ${\bf x}\in \mathbb{F}_2^n$
%{\bf x}
is the set of its non-zero coordinates, that is,
\[ \mbox{supp}({\bf x}) = \{ i\in \{1,2,\ldots ,n\} \mid x_i\neq
0\} , \]
and the weight wt({\bf x}) of {\bf x} is the size $|\supp(\bfx)|$ of its support.

As usual, an $[n,k]$ code $C$ is a $k$-dimensional subspace of
$\mathbb{F}_2^n$; the dual code of $C$, denoted by $C^{\perp}$, is
the $[n,r]$ code with $r=n-k$ consisting of all vectors in $\mathbb{F}_2^n$
that have inner product 0 with all words from $C$. The number $r$ is referred to as
the {\em codimension\/} or {\em redundancy\/} of the code. An $r\times n$
matrix is called a parity check matrix for $C$ if its rows span
$C^{\perp}$. When we speak about ``code'', we will always mean
binary linear code.

A received word containing only correct symbols and erasures can
be decoded unambiguously precisely when exactly one codeword
agrees with this word in the non-erased positions; as we consider
linear codes, this is the case precisely when the erased positions
do not contain the support of a nonzero codeword. This motivates
the following definition.
%LTv7: C-correctable en C-uncorrectable ingevoerd
\begin{defi}\label{uncorr}
Let $C$ be a code of length $n$. A set $E\subseteq\{1,2,\ldots
,n\}$ is called $C$-{\em uncorrectable} if it contains the support
of a non-zero codeword, and $C$-{\em correctable} otherwise.
\end{defi}

 The relevance of the following definition is obvious in
connection with the iterative scheme for erasure decoding
described in the introduction.
\begin{defi}
Let ${\cal H}\subseteq\mathbb{F}_2^n$. A set $E\subseteq\{1,2,\ldots
%HH To avoid overful hbox
,n\}$ is called a {\em stopping set for ${\cal H}$} if
%for each ${\bf h}\in{\cal H}$, we have that wt$({\bf h}(E))\neq 1$.
$\wt(\bfh(E))\neq 1$ for all $\bfh\in\cH$.
\end{defi}
%LTv5 toegevoegd
Note that the empty set is a stopping set as well.
\begin{defi}\label{incorrH}
Let ${\cal H}\subseteq\mathbb{F}_2^n$. A set
$E\subseteq\{1,2,\ldots ,n\}$ is called {\em uncorrectable with
${\cal H}$} if it contains a non-empty stopping set for ${\cal
H}$, and {\em correctable with ${\cal H}$} otherwise.
\end{defi}
%LTv7  with H ipv with respect to H
The iterative correction procedure applied to a set $E$ of
erasures stops with erasures in the largest\footnote{Such a
largest set exists, as the union of stopping sets is again a
stopping set, see \cite{Di02}.} stopping set contained in $E$
\cite[Lemma 1.1]{Di02}. Hence, $E$ is correctable with ${\cal H}$
if and only if the iterative correction procedure, using ${\cal
H}$, removes all erasures. Note that uncorrectable sets with
${\cal H}$ are called {\em dead-end sets} for ${\cal H}$ in
\cite{WeAb05b}.

%LTv7 Zin toegevoegd
Assume we apply the iterative correction procedure with ${\cal H}$
for retrieving words from the code $C$.
%algorithm with a set of parity check equations ${\cal H}$
We are interested in the behavior of the iterative error
correction procedure only for $C$-correctable erasure patterns
(for $C$-uncorrectable erasure patterns, no decoding algorithm can
resolve all erasures). If $C$ has codimension $r$, then for any
$(2^r-r)$ subset~${\cal H}$ of $C^{\perp}$ not containing {\bf 0},
{\em every} correctable erasure pattern is correctable with ${\cal
H}$ \cite[Lemma 8]{WeAb05b}. It is our aim to construct (smaller)
sets of parity check equations ${\cal H}$ such that all
$C$-correctable erasure patterns {\em up to a given cardinality}
are correctable with ${\cal H}$.

 For analysis, the following definition, which deals
with a single step in the iterative decoding algorithm, is useful.

%LT Nieuwe definitie m-erasure reducing
%LTv7 $C$-correctable, "for H" toegevoegd.
\begin{defi}\label{meras}
Let $C$ be a code. An {\em $m$-erasure reducing set for $C$ } is a
subset ${\cal H}$ of $C^{\perp}$ such that no $C$-correctable
erasure pattern of size $m$
%LTv5
%(i.e., no erasure pattern that does not
%contain the support of a nonzero codeword)
is a stopping set for ${\cal H}$.
\end{defi}
%As a consequence, with an $m$-erasure reducing set we can remove,
%for each correctable erasure pattern of size $m$, an erasure with
%one of the parity check equations.
%LTv5 toevoeging
Definition~\ref{meras} has the following consequence. An
$m$-erasure reducing set ${\cal H}$ allows to resolve, for each
$C$-correctable erasure pattern $E$ of size $m$, at least one of
the erasures from $E$ with a parity check equation from ${\cal
H}$.
%LTv5 einde toevoeging
\begin{defi}\label{merascomp}
% Goed, ik heb nu overal "decoding" vervangen door "correcting",
% maar het viel me hier wel op dat de context een iterative
% erasure *decoding* algorithm is; dat pleit juist weer voor het
% woord "decoding" ipv correcting. Blijf je bij je mening, of geeft
% dit je aanleiding te switchen? Ik vind alles best.
% Trouwens de zin
% "for $C$, the iterative decoding scheme can correct all correctable"
% beneden suggereert misschien weer "correcting. Denk er nog een keer over.
Let $C$ be a code. An {\em $m$-erasure correcting set for $C$} is a
 subset ${\cal H}$ of $C^{\perp}$
%such that all dead-end sets for ${\cal H}$
%that have size at most $m$ contain the support of a non-zero codeword.
that is $m'$-erasure reducing for all $m'$ with $1\leq m'\leq m$.
\end{defi}
In other words, with an $m$-erasure correcting set ${\cal H}$ for
$C$, the iterative correction procedure can correct all
$C$-correctable erasure patterns of size {\em at most\/} $m$ by
removing one erasure at the time, without ever getting stuck. The
following example shows that an $m$-erasure reducing set need not
be an $m$-erasure correcting set.
%{\bf Example}
\begin{examp}\label{Ex1}
Let $C$ be the binary $[$5,1,5$]$ repetition code,
      and let
      ${\cal H}$ consist of the four vectors ${\bf h}_1=10001$, {\bf h}$_2=01100$,
      ${\bf h}_3=01111$, and ${\bf h}_4=01010$. Note that $\cH$ spans the dual code
$C^\perp$ of $C$ (which is just the even-weight code of length five). In the table
%HH
      below, we provide for each set of erasures of size four a parity check
      equation that has weight one inside this erasure set.

      \begin{tabular}{cc}
      non-erased position & parity check equation  \\
      1   & ${\bf h}_1$    \\
      2   & ${\bf h}_2$   \\
      3   & ${\bf h}_2$   \\
      4   & ${\bf h}_4$   \\
      5   & ${\bf h}_1$
      \end{tabular}

      The set ${\cal H}$ is therefore 4-erasure reducing for $C$.
      It is, however, not 4-erasure correcting for $C$, as
      $\{2,3,4\}$ is a stopping set that does not
      contain the support of a nonzero codeword.
So for example the erasure set $\{1,2,3,4\}$ is $C$-correctable,
and can be reduced but not corrected by $\cH$.
\end{examp}
Finally, we introduce the notion of a ``generic'' $m$-erasure
reducing and correcting set for codes of a fixed codimension. The
idea is to describe which linear combinations to take given any
full-rank parity check matrix for any such code.
\begin{defi}\label{gendef}
   Let $1\leq m\leq r$. A set ${\cal A}\subseteq \mathbb{F}_2^r$ is
   called {\em generic $(r,m)$-erasure reducing\/} if for any
   $n\geq r$ and for any $r\times n$ binary matrix $H$ of rank~$r$,
    the collection
    $\{{\bf a}H \mid {\bf a}\in {\cal A}\}$
     is $m$-erasure reducing for the code with parity check matrix
     $H$;
% \end{defi}
% \begin{defi} Let $1\leq m\leq r$. A set ${\cal A}\subseteq \mathbb{F}_2^r$ is
the set $\cA$
% it
is called {\em generic $(r,m)$-erasure correcting\/} if it is
generic $(r,m')$-erasure reducing for all $m'$ with $1\leq m'\leq m$.
     \end{defi}
At first sight, Definition~\ref{gendef} seems to be very
restrictive. However, in the next section we will see that
%, in these definitions,
if the linear combinations work for the parity check matrix of the
$r$-th order Hamming code, then they work for {\em any\/} parity
matrix for {\em any\/} code of codimension $r$.
\section{Generic $(r,m)$-erasure reducing and correcting sets}\label{Sec:prop}

%In this section,
Here we will derive several properties of generic $(r,m)$-erasure
reducing and correcting sets. We start with
%LTv7 Nieuw
 a simple and well-known observation.
\begin{lem}\label{fullrank}
Let $H$ be a parity check matrix for a code $C$ of length $n$, and
let $E\subseteq\{1,2,\ldots ,n\}$. The restriction $H(E)$ of the
matrix $H$ has full rank if and only if there is no non-zero word
${\bf c}\in C$ such that supp({\bf c})$\subseteq E$.
\end{lem}
\bpf The matrix $H(E)$ has full rank if and only if no non-empty
subset $I$ of its columns add to {\bf 0}. As $H$ is a parity check
matrix for $C$, the columns indexed by $I$ add to {\bf 0}  if and
only if $I$ is the support of a codeword. \epf The following
characterization of generic $(r,m)$-erasure reducing sets will
often be used.
 \begin{prop}\label{altdef}
   A set ${\cal A}\subseteq\mathbb{F}_2^r$ is generic $(r,m)$-erasure reducing if
and only if for any $r\times m$ matrix $M$ of rank
   $m$ there is a vector {\bf a}$\in$${\cal A}$ such that
   wt({\bf a}$M)=1$.
   \end{prop}
%{\bf Proof.}
\bpf
 First, suppose that ${\cal A}$ is generic $(r,m)$-erasure
reducing.
 Let $M$ be an $r\times m$ matrix with rank $m$.
 Let $H:=(M\mid I)$, where $I$ denotes the $r\times r$ identity matrix,
 and let $C$ denote the code with parity check matrix $H$. As $M$
 has full rank, Lemma~\ref{fullrank} implies that
 the set $E=\{1,2\ldots
 ,m\}$ does not contain the support of a non-zero codeword.
 As ${\cal A}$ is generic $(r,m)$-reducing, there is a vector {\bf a}$\in$${\cal
 A}$ such that $\left({\bf a}H\right)(E) = {\bf a}(H(E))={\bf a}M$
 has weight one.

 Conversely, suppose that ${\cal A}$ is such that for each
 $r\times m$ matrix $M$ of rank $m$ there is a vector {\bf a}$\in$${\cal
 A}$ such that wt$({\bf a}M)=1$. Let $C$ be a code of codimension $r$, and let
$H$ be an $r\times n$ parity check matrix for $C$;
%. Note that
so that $H$ has full rank $r$.
Let $E\subseteq \{1,2,\ldots ,n\}$ have size $m$ and be such
 that it does not contain the support of a nonzero codeword.
 According to Lemma~\ref{fullrank}, the matrix $H(E)$ has rank~$m$,
and hence there is an {\bf a}$\in$${\cal A}$ such that
 ${\bf a}(H(E))=({\bf a}H)(E)$ has weight one.
% $\;\;\;\Box$
\epf
%LTv5: Hamming codes ipv Hamming code
  Hamming codes play a special role: they are
  the "most difficult" codes to create $m$-erasure reducing sets for.
  The following proposition makes this
  statement precise.
\begin{prop}\label{Hamming}
Let $C$ be a $[2^r-1,2^r-r-1]$
  Hamming code, and let $H$ be a parity check matrix for $C$.
  Let $m\leq r$, and let ${\cal A}\subseteq \mathbb{F}_2^r$.
  The set {\cal A} is generic $(r,m)$-erasure reducing if and only if
$\{{\bf a}H\mid {\bf a}\in{\cal A}\}$ is $m$-erasure reducing for $C$.
  \end{prop}
\bpf  This is a direct consequence of Lemma~\ref{fullrank},
  Proposition~\ref{altdef}, and the fact that up to a column
  permutation, each $r\times m$ matrix of rank $m$ occurs in $H$,
  as $H$ contains each non-zero column exactly once.
\epf

\begin{prop}\label{mthenm-1}
% "reducing" door "correcting" vervangen...
Let $2\leq m\leq r$. A generic $(r,m)$-erasure reducing set is a
generic $(r,m-1)$-erasure reducing set.
\end{prop}
\bpf Let ${\cal A}$ be a generic $(r,m)$-erasure-reducing set.
Let $M$ be a binary  $r\times (m-1)$ matrix
of rank $m-1$. We write \[ M= \left[ M_0 \mid {\bf x}^\top \right],
\] where ${\bf x}^\top$ denotes the rightmost column of $M$. Let
${\bf y}^\top$ be a vector in $\mathbb{F}_2^r$ that is not in the
linear span of the columns of $M$, and let $M^{\prime}$ denote the
$r\times m$ matrix defined as
\[ M^{\prime} = \left[ M_0 \mid {\bf y}^\top  \mid {\bf x}^\top+{\bf y}^\top \right] . \]
As $M^{\prime}$ has rank $m$, there exists a vector {\bf a}$\in$${\cal
A}$ such that wt$({\bf a}M^{\prime})=1$. We claim that wt$({\bf
a}M)=1$. This is clear if wt$({\bf a}M_0)=1$, as then ${\bf a}{\bf
x}^\top={\bf a}{\bf y}^\top=0$. If ${\bf a}M_0=0$, then {\bf a}{\bf
y}$^\top=0$ and ${\bf a}({\bf x}^\top+{\bf y}^\top)=1$, or vice versa. In
either case, {\bf a}${\bf x}^\top={\bf a}{\bf y}^\top+{\bf a}({\bf
x}^\top+{\bf y}^\top) = 1$, from which we conclude that in this
case also ${\bf a}M$ has weight 1.
\epf
Note that Proposition~\ref{mthenm-1} implies that the parity check
equations induced by a generic $(r,m)$-erasure reducing set can
also be used to resolve an erasure from a correctable erasure set of size
$m-1,m-2,\ldots$ (we have seen in Example~\ref{Ex1} that this need not hold for
a specific $m$-erasure reducing set for a specific code). In other
words, the following proposition holds.
\begin{prop}\label{red=dec}
Any generic $(r,m)$-erasure reducing set is a generic $(r,m)$-erasure correcting set.
\end{prop}
Note that Proposition~\ref{Hamming} and Proposition~\ref{red=dec}
imply that in \cite{WeAb05a}, Weber and
%HH
  Abdel-Ghaffar in fact construct generic $(3,r)$-erasure
  correcting sets.
%HH: to avoid overfull \hbox
\\
According to Proposition~\ref{red=dec}, the terms ``generic
$(r,m)$-erasure reducing'' and ``generic $(r,m)$-erasure
correcting'' can
% Tenslotte claimt dat dan de sterkste resultaten...
% Ik heb dat nog niet gedaan, maar denk wel dat dat zou moeten daar
% waar we die eigenschap *claimen*, bijvoorbeeld na een constructie.
%HH Dit heb ik geprobeerd door te voeren...
% Dus hierna (bijna) altijd redicing vervangen door correcting.
be used interchangably. In the sequel, we mostly use ``correcting'', and base
%be used interchangably. In the sequel, we use ``correcting'', and base
our results on the characterization given in
Proposition~\ref{altdef}.

Finally, for later reference we explicitly state two simple
results.
\begin{prop}\label{propbasistransform} If ${\cal A}$ is a generic $(r,m)$-erasure
correcting set and if $S$ is any invertible $r\times r$ matrix, then the set $\{
{\bf a}S \mid {\bf a}\in {\cal A}\}$ is generic $(r,m)$-erasure
correcting as well.
\end{prop}
\bpf Let ${\cal A}\subseteq\mathbb{F}_2^r$ be
$(r,m)$-erasure correcting. Let $M$ be an $r\times m$ matrix of rank
$m$. Then the matrix $SM$ is an $r\times m$ matrix of rank $m$ as
well, and so there is a vector {\bf a}$\in$${\cal A}$ such that wt(${\bf
a}(SM)$)=1, so wt((${\bf a}S)M)$=1.
\epf
We will say that two generic $(r,m)$-correcting sets $\cA$ and
$\cB=\{{\bf a}S \mid {\bf a}\in {\cal A}\}$
with $S$ invertible are {\em equivalent\/}.
\begin{prop}\label{span}
%HH Hier reducing laten staan: sterkste resultaat!
For all $r,m$ with $1\leq m\leq r$, a generic $(r,m)$-erasure reducing set spans
$\mathbb{F}_2^r$.
\end{prop}
\bpf Let ${\cal A}\subseteq\mathbb{F}_2^r$ be such that span$({\cal
A})\neq \mathbb{F}_2^r$.  Let {\bf x} be a non-zero vector in
$\left(\mbox{span}({\cal A})\right)^{\perp}$. Let $S$ be any
invertible matrix with ${\bf x}$ as leftmost column.  Finally, let
$M$ be an $r\times m$ matrix of rank $m$ for which the top row has
odd weight and all other rows have even weight. As for  each ${\bf
a}\in{\cal A}$ the vector ${\bf a}S$ starts with a zero, the
vector $({\bf a}S)M$ has even weight.  Consequently, $\{ {\bf
a}S\mid {\bf a}\in{\cal A}\}$ is not a generic $(r,m)$-erasure
reducing set. Now Proposition~\ref{propbasistransform} implies
that ${\cal A}$ is not a generic $(r,m)$-erasure reducing set.
\epf
%LT: Verwijderd! LT
%\begin{prop}
%Let ${\cal A}$ be generic $(r,m)$-erasure reducing. For
%each basis ${\bf b}_1,\ldots ,{\bf b}_m$ of~$\mathbb{F}_2^m$ and
%each $r\times m$ matrix $M$  of rank $m$, there exists a vector ${\bf
%a}\in{\cal A}$ such that ${\bf a}M={\bf b}_j$ for some $j\in
%\{1,\ldots ,m\}$.
%\end{prop}
%\bpf Let ${\bf b}_1,\ldots ,{\bf b}_m$ be a basis for
%$\mathbb{F}_2^m$, and let $M$ be an $r\times m$ matrix of rank
%$m$. Let $B$ be the matrix with rows ${\bf b}_1,\ldots, {\bf
%b}_m$. Clearly, $B$ is invertible, and $MB^{-1}$ has rank $m$.
%Therefore, there exists an ${\bf a}\in {\cal A}$ and a
%$j\in\{1,\ldots ,m\}$ such that ${\bf a}(MB^{-1})= {\bf e}_j$, and
%so ${\bf a}M={\bf e}_jB={\bf b}_j$.
%\epf
%
%
\section{A construction for generic $(r,m)$-erasure correcting sets}\label{Sec:constr}
% HH Hier blijft een overful \hbox (title) zitten ...
%
 We start this section with describing generic $(r,m)$-erasure correcting sets
${\cal A}_{r,m}$ for all $r$ and $m$ with $r\geq m\geq 2$.
We will see that the set ${\cal A}_{r,3}$ is
%related
equivalent to the sets found by Weber and Abdel-Ghaffar
%via an
%element-wise multiplication with an
% invertible $r\times r$ matrix (cf.\ Proposition~\ref{propbasistransform}).
%%
%\subsection{Basic construction}
\begin{theo}
Let $2\leq m\leq r$. The set ${\cal A}_{r,m}$ defined as
\[ {\cal A}_{r,m} = \{ {\bf a}= (a_1,a_2,\ldots ,a_r) \in \mathbb{F}_2^r
\mid a_1=1 \mbox{ and wt}({\bf a})\leq m \}\]
is a generic
$(r,m)$-erasure correcting set of size
\[ \sum_{i=0}^{m-1} {r-1 \choose i} . \]
\end{theo}
\bpf As ${\cal A}_{r,m}$ consists of all vectors that
start with a one and have weight at most $m-1$ in the positions
2,3,\ldots ,$r$, the statement on the size of ${\cal A}_{r,m}$ is
obvious.

In order to show that ${\cal A}_{r,m}$ is indeed generic $(r,m)$-erasure
correcting, we will use Proposition~\ref{altdef}.
So let $M$ be an $r\times m$ matrix of rank $m$. We have to show
that there is a vector {\bf a}$\in$${\cal A}_{r,m}$ such that
wt({\bf a}$M$)=1. To this end, we proceed as follows.
%Proposition~\ref{altdef} implies that we then
%have proved that ${\cal A}_{r,m}$ is generic $(r,m)$-erasure
%correcting. \\
For $1\leq i\leq r$, let {\bf m}$_i$ denote the $i$-th row of $M$.
Let $I\subseteq\{1,2,\ldots ,r\}$ be such that $\{{\bf m}_i \mid
i\in I\}$ forms a basis for $\mathbb{F}_2^m$. We distinguish
two cases.

(i): ${\bf m}_1\neq {\bf 0}$.

\noindent In this case, we can and do choose $I$ such that 1$\in I$. The set $\{
\sum_{i\in I} x_i{\bf m}_i\mid ({x_i})_{i\in I}, x_1=0\}$ is $(m-1)$-dimensional
and hence cannot contain all unit
vectors. So there exists a vector ${\bf x}=(x_i)_{i\in I}$ with
$x_1=1$ and wt($\sum_{i\in I} x_i{\bf m}_i)=1$. Now, let ${\bf
a}\in\mathbb{F}_2^r$ be the vector that agrees with ${\bf x}$ in
the positions indexed by $I$ and has zeroes elsewhere. Then
$a_1=x_1=1$ and wt({\bf a})$=$wt$({\bf x})\leq m$, hence {\bf
a}$\in$${\cal A}_{r,m}$ and ${\bf a}M=\sum_{i=1}^r a_i{\bf m}_i=
\sum_{i\in
I}x_i{\bf m}_i$, so wt$({\bf a}M)=1$.

(ii): ${\bf m}_1 = {\bf 0}$.

\noindent
Note that in this case $1\notin I$.
As $ \{{\bf m}_i\mid i\in I\}$ forms a basis,
%for each unit vector {\bf e}$_j$ there is a vector
there are independent vectors ${\bf x}(j)=(x_i(j)\mid i\in I\}$
such that ${\bf e}_j= \sum_{i\in I} x_i(j){\bf m}_i$ for all $j$.
As there is
just one vector ${\bf x}$ of weight $m$, and there are $m\geq 2$
unit vectors, there is an index $j$ such that wt$({\bf x}(j))\leq m-1$. Now,
let {\bf a} be the vector that agrees with ${\bf x}(j)$ in the
positions indexed by $I$, has a ``1'' in the first position, and zeroes
elsewhere. As wt$({\bf x}(j))\leq m-1$, the vector {\bf a} is in
${\cal A}_{r,m}$. Moreover, we have that ${\bf a}M = \sum_{i=1}^n
a_i{\bf m}_i = a_1{\bf m}_1 + \sum_{i\in I}a_i{\bf m}_i= {\bf 0} +
{\bf e}_j={\bf e}_j$.
\epf
We now compare our result for $m=3$ with that of Weber and
Abdel-Ghaffar \cite{WeAb05a}, which in our terminology states that
\[ {\cal W}_r = \{ {\bf e}_i \mid 1\leq i\leq r\} \cup
                  \{ {\bf e}_1 + {\bf e}_i + {\bf e}_j\mid
                     2\leq i < j \leq r\} \]
                     is generic $(r,3)$-erasure correcting.
To this end, let $S$ be the matrix with the all-one vector as
leftmost column, and with ${\bf e}_j^\top$ as $j$-th column for $2\leq j\leq r$.
Obviously $S$ is invertible, and
\[ {\bf e}_1S={\bf e}_1, \qquad
\mbox{$({\bf e}_1+{\bf e}_i)S= {\bf e}_i$}, \qquad
\mbox{$({\bf e}_1+{\bf e}_j+{\bf e}_k)S = {\bf e}_1+{\bf e}_i+{\bf e}_j$} \]
for $2\leq i\leq r$ and $2\leq j< k \leq r$.
As a consequence, we have that
\[ {\cal W}_r = \{ {\bf a}S \mid {\bf a}\in {\cal A}_{r,3}\} .
\]
So ${\cal W}_r$ and ${\cal A}_{r,3}$ are related via an
element-wise  multiplication with an invertible matrix,
hence they are equivalent (see Proposition~\ref{propbasistransform}).
%So ${\cal W}_r$ and ${\cal A}_{r,3}$ are related via an
%element-wise  multiplication with an invertible matrix, which
%according to Proposition~\ref{propbasistransform} maps generic
%$(r,m)$-erasure correcting sets to generic $(r,m)$-erasure
%correcting sets.
%
%
\section{Some optimality results}\label{Sec:optimal}
%
%In Subsection~\ref{optimal},
%Here, we first show that the set $A_{r,m}$  is an
%$(r,m)$-erasure correcting sets of minimal size for $m=2$ and $m=r$.
%\footnote{We in fact characterize the generic $(r,r)$
%erasure correcting sets of minimum size}
%We will even characterize generic $(r,r)$-erasure correcting sets of minimum size.
%In Subsection~\ref{negative},
In this section we investigate the minimum size $F(r,m)$
of a generic $(r,m)$-erasure correcting set, where $1\leq m\leq r$.
We first show that $F(r,1)=r$ and that for $m=2$ and $m=r$, the set ${\cal
A}_{r,m}$ is a generic $(r,m)$-erasure correcting set of minimal
size. Moreover, we also characterize all generic $(r,r)$-erasure
correcting sets of minimum size.
\begin{prop}
We have that $F(r,1)=r$ for $r\geq1$ and $F(r,2)=r$ for $r\geq2$.
\end{prop}
\bpf The case $r=1$ is trivial; for $r\geq2$ the proposition is a
direct consequence of Proposition~\ref{span} (for the lower bound)
% te doen...
and the fact that $\cA_{r,2}$ is generic $(r,2)$-erasure correcting
of size $r$.
%hence also generic $(r,1)$-erasure by Proposition~\ref{mthenm-1}.
\epf
\begin{theo}
If ${\cal A}\subseteq \mathbb{F}_{2^r}$ is a generic $(r,r)$-erasure
correcting set, then $|{\cal A}| \geq 2^{r-1}$. Equality holds if
and only if $\mathbb{F}_2^r\setminus {\cal A}$ is a hyperplane,
i.e. an $(r-1)$ dimensional subspace of $\mathbb{F}_2^r$;
so up to equivalence the unique optimal set is $\cA_{r,r}$.
As a consequence, $F(r,r)=2^{r-1}$.
\end{theo}
\bpf
Let $\cA\subseteq \mathbb{F}_{2^r}$ be generic $(r,r)$-erasure correcting.
We claim that the complement $\mathbb{F}_{2^r}\setminus \cA$ does not
contain $r$ independent vectors. Indeed, let
${\bf u}_1,{\bf u}_2,\ldots ,{\bf u}_r$ be
independent vectors. Let $U$ be the matrix with ${\bf u}_i$ as
$i$-th row, and let $M:=U^{-1}$. For 1$\leq j\leq m$, ${\bf
a}M={\bf e}_j$ if and only if ${\bf a}={\bf e}_jM^{-1} = {\bf
e}_jU={\bf u}_j$. As ${\cal A}$ is generic $(r,r)$-erasure
correcting, at least one of the ${\bf u}_j$'s is indeed in ${\cal A}$.

So the complement of ${\cal A}$ does not contain
$r$ independent vectors, and hence lies in a subspace of dimension $r-1$;
we conclude that
%\[ 2^r - \mid {\cal A}\mid \;  = \; \mid \mathbb{F}_2^r\setminus {\cal A} \; \mid
%   \leq 2^{r-1}, \]
\[ \mid {\cal A}\mid \;  = \; 2^r-\mid \mathbb{F}_2^r\setminus {\cal A}
\mid \;\;
   \geq 2^{r-1}, \]
%  In case of equality, $\mathbb{F}_2^r\setminus{\cal A}$, being a
%  set with $2^{r-1}$ elements that does not contain $r$
%  independent vectors, is a hyperplane.  \\
with equality if and only if the complement $\mathbb{F}_2^r\setminus{\cal A}$
is a hyperplane in $\mathbb{F}_2^r$.

Conversely, suppose that $\mathbb{F}_2^r\setminus{\cal A}$ is a hyperplane.
Then a basis $\{{\bf u}_1,{\bf u}_2,\ldots ,{\bf u}_r\}$ for $\mathbb{F}_2^r$ can be
found such
  that $\{{\bf u_2},{\bf u}_3,\ldots ,{\bf u}_r\}$ spans
  $\mathbb{F}_2^r\setminus {\cal A}$, and so
  \[ {\cal A} = \{ \sum_{i=1}^r a_i{\bf u}_i \mid
                        (a_1,a_2,\ldots ,a_r)\in\mathbb{F}_2^r
                        \mbox{ and } a_1 = 1 \} . \]
   Let $U$ be the $r\times r$ matrix with ${\bf u}_i$ as $i$-th
   row. For each ${\bf x}=(x_1,x_2,\ldots ,x_r)\in F_2^r$, we have that
  \[  {\bf x}U =
       (\sum_{i=1}^r x_i{\bf e}_i)U =
         \sum_{i=1}^r x_i({\bf e}_iU) =
         \sum_{i=1}^r x_i{\bf u}_i, \]
 and so we have that
 \[ {\cal A} = \{ {\bf x}U \mid {\bf x}=(x_1,x_2,\ldots x_r)\in
 \mathbb{F}_2^r \mbox{ and } x_1 = 0 \} =
    \{ {\bf a}U \mid {\bf a}\in {\cal A}_{r,r}\} . \]
   As ${\cal A}_{r,r}$ is generic $(r,r)$-erasure correcting,
   and $U$ is invertible, Proposition~\ref{propbasistransform} implies that ${\cal A}$
is also generic $(r,r)$-erasure correcting. \epf

Next we investigate the inclusion-minimality of the sets $\cA_{r,m}$.
First we show that removal of any word of
weight $m$ or $m-1$  from ${\cal A}_{r,m}$ results in a set that
is no longer generic $(r,m)$-erasure correcting; we also show that
removing {\em any} word from  ${\cal A}_{r,r-1}$ results in a set
that no longer is generic $(r,r-1)$-erasure correcting. Finally,
%in Section~\ref{positive},
we show that if $r\geq 2^{m-1}+1$, then certain
%subsets of ${\cal A}_{r,m}$
words of weight less than $m-2$ can be removed such that the
resulting set still is
%while keeping a generic
generic $(r,m)$-erasure correcting.
% set if $r\geq 2^{m-1}+1$.
%\subsection{Some words cannot be removed from ${\cal
%A}_{r,m}$}\label{negative}
% In this section, we show that if we
%remove any word of weight $m-1$ or $m$ from ${\cal A}_{r,m}$, the
%resulting set is no longer generic $(r,m)$-erasure correcting. We
%also show that the same is true for {\em any} word from ${\cal A}_{m+1,m}$.

%SHOULD THIS BE REALLY IN? I PROPOSE THAT PERHAPS WE SIMPLE CLAIM THIS RESULT.
\begin{prop}\label{cannotremove}
Let $r\geq m\geq 3$. If ${\bf a}\in{\cal A}_{r,m}$ has weight $m$
or $m-1$, then ${\cal A}_{r,m}\setminus \{{\bf a}\}$ is not a
generic $(r,m)$-erasure correcting set.
\end{prop}
\bpf Let ${\bf a}\in{\cal A}_{r,m}$ have weight at least
$m-1$. We will construct an $r\times m$ matrix $M$ such that ${\bf
a}$ is the only vector {\bf x} in ${\cal A}_{r,m}$ such that
wt({\bf x}$M$)=1.

First, assume that ${\bf a}$ has weight $m$. We assume without
loss of generality\footnote{If not, we can transform {\bf a} to
(1,\ldots ,1,0,\ldots 0) by a coordinate permutation that fixes 1,
and apply the same permutation to the rows of the  matrix found
below.} that ${\bf a}=(1,1,\ldots,1,0,\ldots,0)$. Let $M$ be the
$r\times m$ matrix defined as
\[ M = \left( \matrix{ 1 & 1\ldots 1 \cr
                0 & I_{m-1}    \cr
                 0 & 0\ldots 0  \cr
                 \vdots & \vdots  \cr
                 0 & 0\ldots 0} \right), \]
 where $I_{m-1}$ denotes the identity matrix of order $m-1$.
 It is clear that $M$ has rank $m$.
Now let ${\bf x}=(x_1,x_2,\ldots ,x_{r})\in {\cal A}_{r,m}$. As
${\bf x}_1 = 1$, we have that
\[ {\bf x}M = (1,1+x_2 ,\ldots ,1+x_m) . \]
Consequently, if ${\bf x}M$ has weight 1, then $x_1=x_2=\ldots
=x_{m}=1$, so ${\bf a}$ is the only vector {\bf x}
in ${\cal A}_{r,m}$ for which ${\bf x}M$ has weight 1.

Next, assume that {\bf a} has weight $m-1$; we assume without loss
of generality that {\bf a} starts with $m-1$ ones. Let $M$ be the
$r\times m$ matrix defined as
\[ M = \left( \matrix{ 1 & 1 \ldots 1 & 0 \cr
                       0 & I_{m-2}    & 0 \cr
                       0 & 0\ldots 0 & 1 \cr
                       \vdots & \vdots & \vdots \cr
                       0 & 0 \ldots 0 & 1} \right) . \]
Clearly, $M$ has rank $m$. Now let ${\bf x}=(x_1,x_2,\ldots,
x_r)\in{\cal A}_{r,m}$. As ${\bf x}_1=1$, we have that
\[ {\bf x}M = (1,1+x_2,\ldots ,1+x_{m-2},1+x_{m-1}, x_m + \ldots + x_r). \]
Hence, if ${\bf x}M$ has weight 1, then $x_2=x_3=\ldots
=x_{m-1}=1$, and $x_m + \ldots +x_{r}=0$. As {\bf x} is in
$A_{m,r}$, it has weight at most $m$; as {\bf x} start with $m-1$
ones, and has an even number of ones in the positions
$m,m+1,\ldots ,r$, it follows that $x_m=x_{m+1}=\ldots =x_{r}=0$,
and so ${\bf x}={\bf a}$.
\epf
%SO PERHAPS CONTINUING HERE...

%BUT I WOULD LIKE TO KEEP THIS IN, AND ADD THE CONJECTURE...
\begin{prop} Let $m\geq 3$. No subset of ${\cal A}_{m+1,m}$ is
generic $(m+1,m)$-erasure correcting.
\end{prop}
\bpf Let {\bf a} be a vector of weight $w$, 1$\leq w\leq
m$, in ${\cal A}_{m,m+1}$. We show that ${\cal A}_{m,m+1}\setminus
\{ {\bf a}\}$ is not generic $(m+1,m)$-erasure correcting by
constructing an $(m+1)\times m$ matrix $M$ such that ${\bf a}$ is
the only vector {\bf x} in ${\cal A}_{m+1,m}$ such that {\bf x}$M$
has weight 1. We assume without loss of generality that ${\bf a}$
starts with $w$ zeroes. Let $M$ be the matrix
\[ M = \left( \matrix{ 1 & 1 \ldots 1 & 0 \ldots 0 \cr
                       0 &  I_{w-1}   & 0 \ldots 0 \cr
                       0 & 0\ldots 0  & I_{m-w} \cr
                       0 & 0\ldots 0  & 1 \ldots 1} \right) . \]
Clearly, $M$ has rank $m$. Let ${\bf x}=(x_1,x_2,\ldots ,x_{m+1})
\in {\cal A}_{m+1,m}$.   As $x_1=1$, we have that
\[ {\bf x}M = (1,1+x_2,\ldots ,1+x_w,x_{w+1}+x_{m+1},\ldots
,x_m + x_{m+1}) . \] Hence, if wt(${\bf x}M)=1$, then $x_j=1$ for
$1\leq j\leq w$, and $x_{w+1}=x_{w+2}=\ldots =x_{m+1}$. As
wt$({\bf x})\leq m$, it follows that $x_{j}=0$ for $j\geq w+1$,
and so ${\bf x}={\bf a}$. \epf In fact, although we cannot prove
it yet, we have reason to believe that the following is true.
\begin{con}\label{optr-1} If $r\geq 2$, then $A_{r,r-1}$
is the smallest possible
$(r,r-1)$-erasure correcting set, that is, $F(r,r-1)=2^{r-1}-1$.
\end{con}

%\subsection{Subsets of ${\cal A}_{r,m}$ that are generic $(r,m)$-erasure
%correcting}\label{positive}
%Theorem~\ref{cannotremove} shows that ${\cal A}_{r,m}$ cannot be
%transformed to a smaller generic $(r,m)$-erasure correcting set by
%removing words of weight $m$ or $m-1$. The next theorem shows that
%if $r$ is sufficiently large, some words of weight at most $m-2$
%can be removed from ${\cal A}_{r,m}$ such that the remaining set
%still is generic $(r,m)$-erasure correcting.

We now come to one of the main results stating that the sets
$\cA_{r,m}$ are not optimal if $m\geq3$ and $r$ is large with
respect to $m$. The precise statement is as follows.
\begin{theo}\label{imprgeneral}
Let $r\geq m$ and $r\geq 2^{m-1}+1$.  Let
${\cal B}_{r,m}$ be defined as
\[ \cB_{r,m} = \{ {\bf a}\in {\cal A}_{r,m} \mid
    \mbox{wt}({\bf a}) \leq m-2 \mbox{ and }
               \mbox{supp}({\bf a})\subseteq\{1,2\ldots ,r-2^{m-1} \}
               \} . \]
Then
\[{\cal B}_{r,m}= \sum_{i=0}^{m-3} {r-2^{m-1}-1 \choose i}\]
and
$\cA^*_{r,m}:={\cal A}_{r,m}\setminus{\cal B}_{r,m}$ is a generic
$(r,m)$-erasure correcting set.
\end{theo}
\bpf Let $M$ be an $r\times m$ matrix of rank $m$. We
will show that there is an {\bf x}$\in {\cal A}_{r,m}\setminus
{\cal B}_{r,m}$ such that wt$({\bf x}M) = 1 $. We denote the set
$\{r-2^{m-1}+1,\ldots ,r\}$ by $I$,  and the $i$-th row of $M$ by
${\bf m}_i$. \\
 As ${\cal A}_{r,m}$ is generic
$(r,m)$-erasure correcting, there is an ${\bf a}\in{\cal A}_{r,m}$
such that wt$({\bf a}M)=1$, say {\bf a}$M = {\bf e}_1$. Let us
assume that ${\bf a}\in {\cal B}_{r,m}$, as otherwise we can take
${\bf x}={\bf a}$. We will add to {\bf a} a vector of weight 1 and
or 2 with support in $J$ such that the resulting vector {\bf x},
which is automatically in ${\cal A}_{r,m}\setminus {\cal
B}_{r,m}$, satisfies wt$({\bf x}M) = 1$. We distinguish four
cases.

(i): For some $i\in I$ and some $j\in\{1,2,\ldots ,m\}$, we have
that ${\bf m}_i= {\bf e}_{1}+{\bf e}_j$. Then we take ${\bf
x}={\bf a}+{\bf e}_i$; note that  {\bf x}$M={\bf e}_1+{\bf
m}_i={\bf e}_j$.

(ii): For some distinct $i,j\in I$,  ${\bf m}_i={\bf m}_j$. Then
we take ${\bf x}={\bf a}+{\bf e}_i+{\bf e}_j$.

(iii): For some $i,j\in I$ and some $k\in \{2,3,\ldots,m\}$, ${\bf
m}_i={\bf e}_1$ and ${\bf m}_j={\bf e}_k$. Then we take ${\bf
x}={\bf a}+{\bf e}_i+{\bf e}_j$.

(iv): Finally, assume that we are in neither of the above cases.
Let $V:= \{ {\bf m}_i\mid i\in I\}$.  Because we are not in case
(b), $|V|=|I|=2^{m-1}$.
 For ${\bf y}=(y_1,\ldots ,y_{m-1})\in\mathbb{F}_2^{m-1}$,
 let $T({\bf y}):=\{(0,y_1,\ldots ,y_{m-1}),(1,y_1,\ldots ,y_{m-1})\}$.
 As we are not in case a, $(T({\bf 0})\cap V)\subseteq\{{\bf e_1}\}$, and
 for $i=1,2,\ldots ,m-1$, $(T({\bf e}_i)\cap V)\subseteq\{{\bf e}_{i+1}\}$.
 As we are not in case c, $V$ contains at most $m-1$ unit vectors,
 and so
 \[ \sum_{y\in\mathbb{F}_2^{m-1}\mid \mbox{{\small wt}}({\bf y})\leq
 1} |T({\bf y})\cap V| \leq m-1 . \]
 As a consequence, we have that
\begin{equation}\label{setsize}
  \sum_{y\in\mathbb{F}_2^{m-1}\mid \mbox{\small{wt}}({\bf y})\geq
  2} |T({\bf y})\cap V| \;\; \geq \; |V| -(m-1) = 2^{m-1}-m+1.
     \end{equation}
As there are $2^{m-1}-m$ vectors in $\mathbb{F}_2^{m-1}$ of weight
at least $2$, Equation~\ref{setsize} implies that there is a
vector ${\bf y}$ of weight at least 2 such that $|T({\bf y})\cap
V| = 2$. That is, there are row indices $i$ and $j$ in $I$ such
that ${\bf m}_i=(0{\bf y})$ and ${\bf m}_j=(1{\bf y})$. We take
${\bf x}={\bf a}+{\bf e}_i+{\bf e}_j$.
\epf

In fact it is not too difficult to show that the lower bound $2^{m-1}+1$ on $r$ in
Theorem~\ref{imprgeneral} is optimal,
in the sense that $\cA^*_{r,m}$ is not generic $(r,m)$-erasure correcting
if $r=2^{m-1}$.

Theorem~\ref{imprgeneral} has the following interesting
consequence.
\begin{cor} For
$r\geq 5$, the set $A_{r,3}\setminus\{ {\bf e}_1\}$ is generic
$(r,3)$-erasure correcting. \end{cor}
%LT: De auteurs vragen zich dat daar niet af.
%This result answers the question in \cite{WeAb05a} about the optimality of their
%construction: it is not.
The construction from \cite{WeAb05a} therefore is not optimal
(although the improvement of course only is marginal).
%LT Einde toevoeging
%
%
\section{Concluding remarks}\label{Sec:claims}
%
%
%HH Hier komt reducing weer voor
In this paper, we introduced and studied generic $(r,m)$-erasure
reducing and correcting sets.
% and generic $(r,m)$-erasure decoding sets.
An obvious extension of this work is to consider $(r,m)$-erasure reducing or
%HH
correcting
sets that are  generic for a certain {\em class\/} of codes of codimension $r$ only.
%parity check matrices.
As an example, let us consider
%an $r\times n$ parity check matrix $H$ of an
the class of {\em even weight\/} codes.
Such codes have an $r \times n$ parity check matrix $H$
for which the first row consists of the all-one vector.
Let $E\subseteq\{1,2,\ldots ,n\}$ have size $m=4$
and suppose that $H(E)$ has rank four. We claim that there is an
$i\in\{2,\ldots ,r\}$ such that ${\bf e}_iH(E)$ or $({\bf
e}_1+{\bf e}_i)H(E)$ has weight one. Indeed, as $H(E)$ has rank four,
it contains a row of odd weight, say its $i$-th row. If this row
has weight one, then ${\bf e}_iH(E)$ has weight one; if not, this
row has weight three, and so $({\bf e}_1+{\bf e}_i)H(E)$ has weight
one. As a consequence,
%we can manage with $2(r-1)$ vectors,
we have a set of size $2(r-1)$ that is generic $(r,4)$-erasure
%HH
reducing (in fact even generic $(r,4)$-erasure
correcting) {\em for even-weight codes\/}.
%while the set ${\cal A}_{r,4}$ has size of the order $r^3$!
(Note that the set ${\cal A}_{r,4}$ has a size of the order $r^3$.)
%as many as $\sum_{i=0}^{3}{r-1 \choose i}$ elements!
A manuscript on generic $(r,m)$-erasure reducing and correcting
sets for even weight codes is in preparation.

%All sets ${\cal A}_{r,m}$ are contained in the set
%$\{(x_1,x_2,\ldots, x_r)\in\mathbb{F}_2^r\mid x_1=1\}$. The
%authors presently are studying more sophisticated constructions of
%generic $(r,m)$-erasure correcting sets contained in that set. The
%results obtained thus far are promising.
%For example, we can construct generic $(r,3)$-erasure correcting
%sets of a size that for large $r$ behaves like $r^{\log_2(3)}$,
%which is a big improvement over the size ${r\choose 2}$ of
%$A_{r,3}$. Also, with the theory developed
%so far,
%we obtain generic $(2s,2s-2)$-erasure correcting sets of size at
%most $2^{2s}-2^s$, which significantly improves over the size
%$2^{2s}-2s$ of ${\cal A}_{2s,2s-2}$. These and possible further
%constructions will be reported in a follow-up paper. \\
%In this follow-up paper, we also show by an averaging argument
%that for each $m\geq 1$, there exists a constant $c_m$ such that
%for each $r\geq m$, there is an $(r,m)$-erasure correcting set of
%size $c_m\cdot r$.

%HH
We believe that the exact determination of $F(r,m)$ is a difficult problem in
general. However some progress may be possible for the case where $m$ is close to $r$.
Also, it is interesting to study $F(r,m)$ for fixed $m$ and large $r$.
Both these cases will be the subject of a follow-up paper.
% Of toch expliciet dit:
%In a follow-up paper we will show that for fixed $m$ in fact $F(r,m)$ is of
%{\em linear\/} order in $r$.

%
%

\end{document}